\documentclass[12pt,preprint]{aastex}
\usepackage{graphicx}
\begin{document}

\title{How certain is the distance to the most luminous supernova?}

\author{H. Arp}
\affil{Max-Planck-Institut f\"ur Astrophysik, D-85741 Garching, Germany}
\email{arp@mpa-garching.mpg.de}

\begin{abstract}

A recent supernova has been reported as exceeding ``the light output
of an ordinary supernova by at least two orders of magnitude''. It is
noted that it falls in a minor galaxy in the Perseus Cluster. Some
evidence indicating a ten times closer distance for the Perseus
Cluster than its redshift distance is discussed here.

\end{abstract}

\section{Introduction}

In an informative display of supernovae light curves (Fig.1 in Physics
Today, July 2007, p 17) it was clear that SN2006gy was about ``10 times
brighter than the peak luminosity of type Ia''. Implied was a total
radiated energy ``two orders of magnitude'' greater than ordinary
supernovae.

The first question that naturally arises is: With what certainty is its 
distance from the observer known? The Smith et al. paper$^1$ mentions  
that the galaxy in which SN2006 appears is a ``minor member of the
Perseus cluster''. Indeed it is, NGC 1260, with a redshift of 5703
km/sec. But readers who are familiar with the sky recognize this as
the Perseus - Pisces cluster which extends over large regions of the
sky. In fact it extends in filaments over about 90 degrees in angle
which would require a structure of startlingly large size at its
redshift distance of 74 Mega parsecs.
       
\section{The Perseus-Pisces Cluster}

Detailed information on this large region is available from the long
term Cataloguing work done by Fritz Zwicky and his associates. The
galaxies in this large region down to the classification limit of the
Palomar 18-inch Schmidt are shown in Arp$^2$ Figs. 13 - 15. The strongest
line of galaxies, however, consists mostly of E's and S0's originating from
the large Sb (NGC 891) and ending on NGC 1260 and the supernova SN2006gy. 
All the galaxies with redshifts $4800 < cz < 6000$ as given in NED are
shown here in Figure 1.

\begin{figure}[h]
\includegraphics[width=15.0cm]{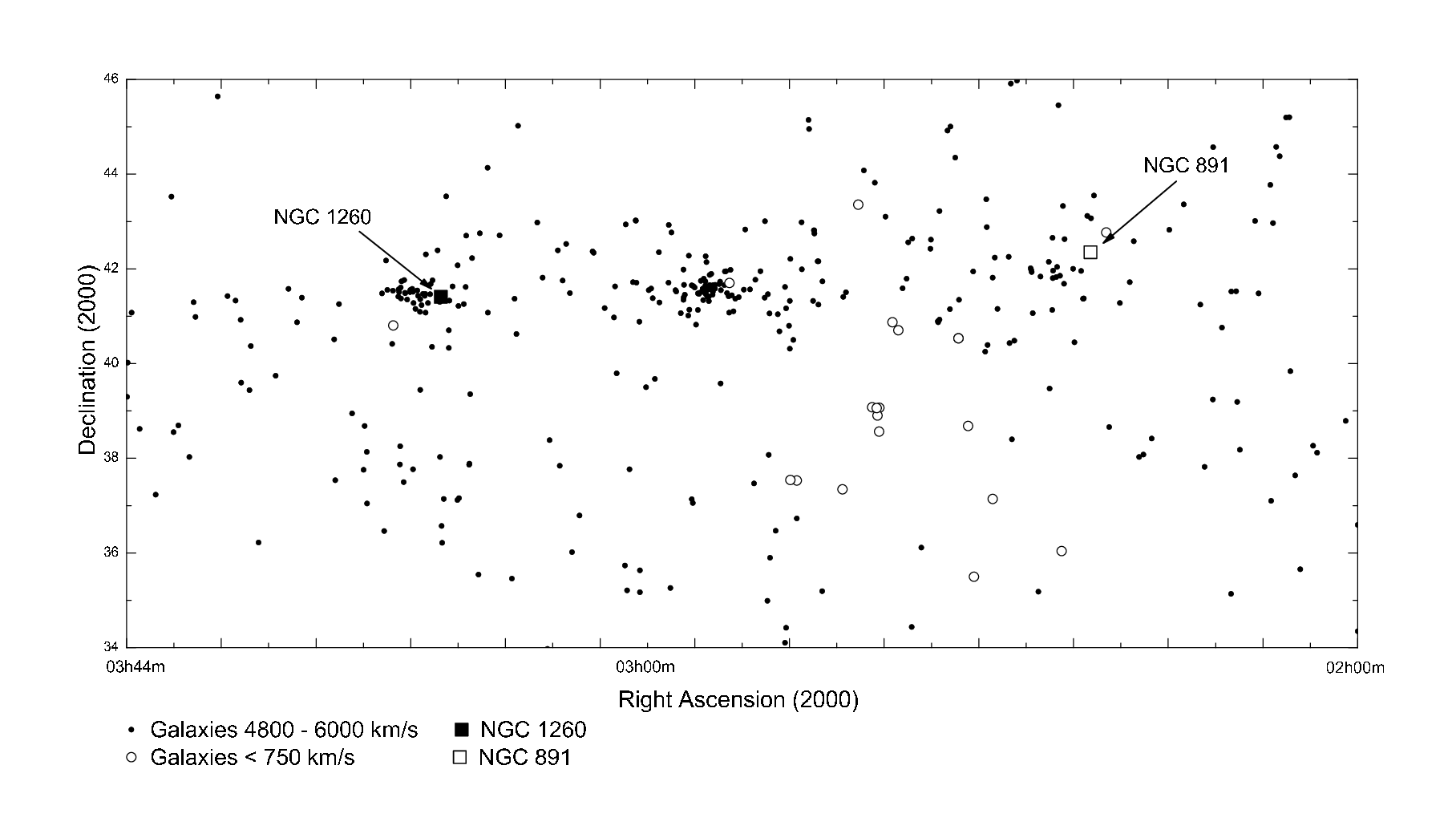}
\caption{Computer plot of all galaxies $4800 < cz < 6000$ in the
eastern end of the Perseus-Pisces filament (from the NED Catalogue
courtesy D. Carosati). Note that the subclusters are elongated along
the filament and the filament appears to start at NGC 891.
\label{fig1}}
\end{figure}

Although the galaxies in the line originating in NGC 891 are
predominantly in the 5300 to 5700 km/sec range of redshift, the
redshift of NGC891 itself is only cz = 528 km/sec. It is intriguing to 
note that it would imply a distance closer by a factor 10 and
luminosity smaller by about a factor of 100 for the supernova if its
progenitor had been ejected out of NGC 891 along with some of the
higher redshift material.

In fact there are 13 smaller galaxies within 5 deg of NGC 891 with
redshifts $527 < v_0 < 637 km/sec$. One could take the least radical 
position that SN2006gy ocurred in such a companion that was faint enough
to escape spectroscopic measurement. The near proximity (~1'') to NGC 1260
would then be an accident. But I believe there is enough evidence that
large galaxies like NGC 891 eject material which becomes higher
redshift companions that I would argue for the ejection of the
supernova progenitor in that process. 

\section{Does the supernova progenitor come from NGC 891?}

That NGC 891 could have ejected the extremely active radio, X-ray,
infra red galaxy Perseus A (NGC 1275) is suggested by the
elongation of the perseus A cluster back along the line to NGC
891. Moreover the NGC 891 minor axis is only about 18 deg. off the
line to Per A. Of course the low redshift material which gave rise to
the supernova would have had to have been entrained along with the
ejection of the higher redshift cluster galaxies.

That companion galaxies have an ejection origin along the minor axes
of edge on disk galaxies was first pointed out by Holmberg$^3$ whose  
observations demonstrated in 1969 preferential alignment along minor
axes of disk galaxies. It was then demonstrated by Arp$^4$ and
L\'opez Corredoira and Guti\'errez$^5$ that companions came out in a
cone with half opening angle about 35 degrees and quasars within $\pm20$ 
degrees. As mentioned, the minor axis of NGC 891 is aligned with the
nearest galaxies in Fig 1 and to within about 18 deg. of the Perseus
cluster to the east.

As for entraining low redshift material from the parent there are now
a number of cases where smaller dwarfs are considered to be physical  
companions which have arisen as fragments from disturbed regions of
parent galaxies. A good example of this is NGC 5985 in Fig.2 where
this Seyfert galaxy has ejected a quasar but only 2.4 arcsec from it
is attached a dwarf with the same small redshift as the big, low
redshift, parent galaxy.

\begin{figure}[ht]
\includegraphics[width=14.0cm]{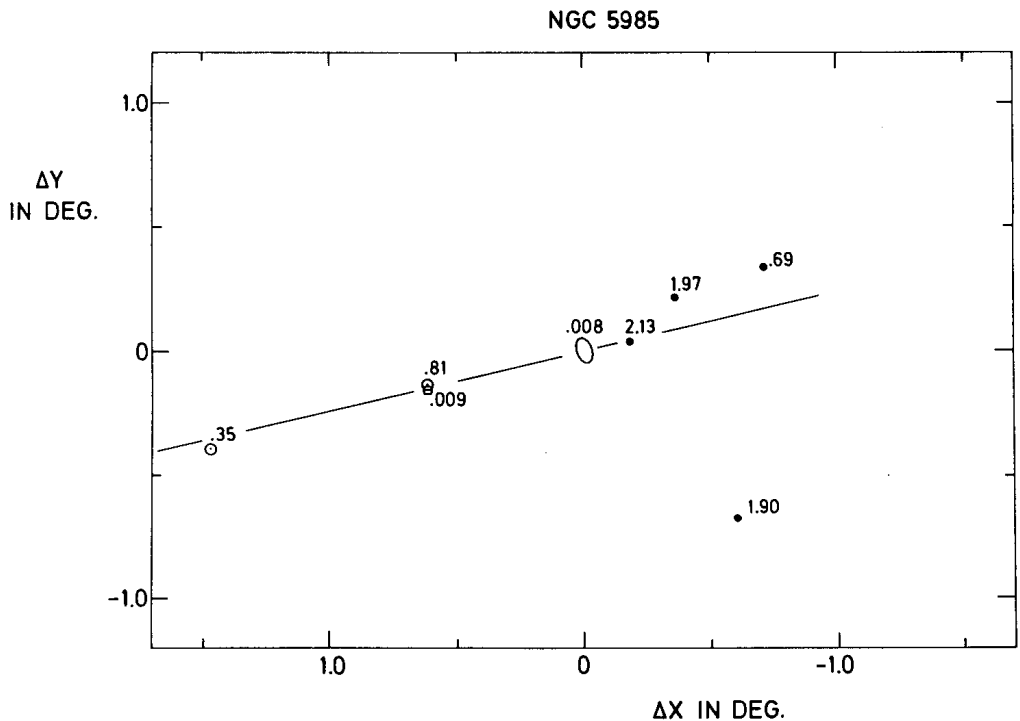}
\caption{The minor axis of the Seyfert Galaxy NGC 5985 (z = .008)
leads through the quasar (z = .81) and the dwarf galaxy (z = .009).
The latter two are separated by 2.4''.
\label{fig2}}
\end{figure}

\section{Lines of higher redshift galaxies from bright apparent
magnitude galaxies}. 

But perhaps the strongest argument for the ejection origin of
the Perseus Cluster extending eastward from NGC 891 is the study of
the 20 brightest galaxies in apparent magnitude north of Declination = 0
deg. Of the 14 that are uncrowded by nearby bright galaxies, a total
of 13 have well marked lines and concentrations of fainter, higher
redshift galaxies$^2$. Figure 3 shows the range in redshift of the
lines of fainter galaxies which are strongly associated with almost
every bright galaxy in the sky. This 1990 study was followed by a
2001 study of 14 examples of higher redshift Abell clusters which were
paired and aligned across bright apparent magnitude galaxies$^6$.

\begin{figure}[ht]
\includegraphics[width=13.0cm]{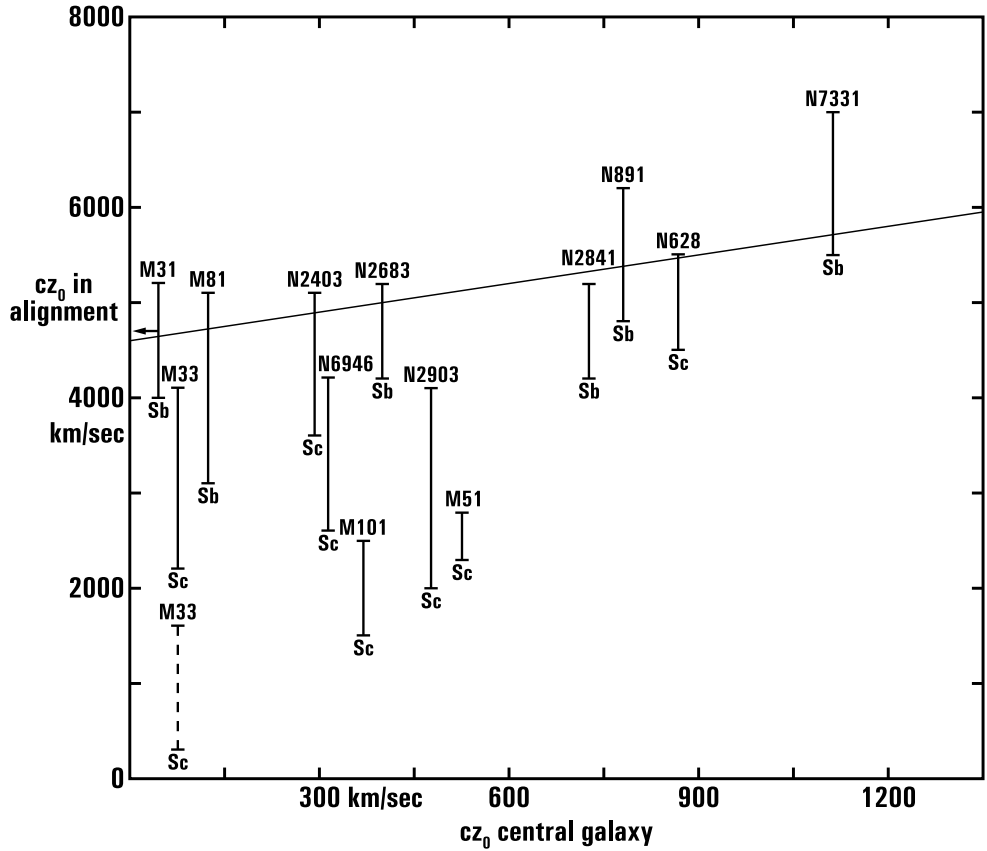}
\caption{Redshift ranges of galaxies which are aligned with
bright apparent magnitude spirals (ordinate) correlated with
redshift of the bright apparent magnitude galaxy (abscissa).
\label{fig3}}
\end{figure}

If such strong ejections take place it is reasonable to expect some of
the material of the ejecting galaxy to be carried along with it. Along
with ejected gas and plasma which is observed we only need to include 
a few old stars or groups of stars among which to find the very
occasional supernova.

Supernovae of type Ia are much discussed today because of their
interpretation in terms of a universe that is expanding faster now
than in the past - in other words, ``dark energy'.' But if their
luminosity is over estimated the sole reason for this radical
postulate falls. If the distance to SN2006gy is strongly
overestimated this may also effect to some extent the models of 
supernovae on which dark energy has been hypothesized.  

Generally one could reason that objects at appreciably large
distances tend to be younger because of the look-back time. Younger
objects tend have built up less metal abundance which in turn lowers
the luminosity of objects like Cepheid variables. If intrinsic
redshifts of younger material is operative, galaxies like the Perseus
Cluster would need low particle masses$^4$ to explain their apparent
association with NGC 891. The smaller elementary particle masses would
give smaller luminosities of objects in those galaxies. All of these
effects would combine in the direction of making the Hubble constant
appear lower in the past and lead to the impression that it was
speeding up at present.

Ironically, if we move SN2006gy to this closer distance it could
reinstate supernovae I as, at least, approximate distance indicators.

\section{References}

1. N. Smith et al. Astrophys. J. and
   http://arXiv.org/abs/astro-ph/0612617 (version 3).

2. H. Arp, J. Astrophys. Astr. (India) {\bf 11}, 411 (1990)

3. E. Holmberg, Ark. Astron. {\bf 5}, 305 (1969).

4. H. Arp, Astrophys. J. {\bf 496}, 661 (1998).
   H. Arp, Seeing Red (pp.86-87) Apeiron, Montreal. 

5. M. L\'opez and C.M. Guti\'errez, Astron. Astrophys. {\bf 461}
   59. (2007) 

6. H. Arp and D. Russell, Astrophys. J. {\bf 549}, 802 (2001)

\end{document}